\documentstyle[preprint,aps]{revtex}
\begin{document}

\draft

\preprint{$
\begin{array}{l}
\mbox{UCD--95--17}\\[-3mm]
\mbox{UCLA/95/TEP/24}\\[-3mm]
\mbox{AMES-HET-95-01}\\[-3mm]
\mbox{FERMILAB-PUB-95/160-T}\\[-3mm]
\mbox{June 1995}\\
\end{array}
$}
\title{Top-Quark Decay Via Flavor-Changing Neutral Currents
\\ at Hadron Colliders}

\author{T. Han$^a$, R.D. Peccei$^b$, and X. Zhang$^c$ }

\address{{ $^a$Department of Physics, University of California, Davis,
 CA 95616, USA}\\
{$^b$Department of Physics, University of California, Los Angeles,
 CA 90024, USA}\\
{$^c$Department of Physics and Astronomy,
 Iowa State University, Ames, IA 50011, USA}\\
{ and  }\\
{Fermi National Accelerator Laboratory, P. O. Box 500,
Batavia,  IL 60510, USA}}

\maketitle

\begin{abstract}

We study low
energy experimental constraints on an
anomalous top-quark coupling associated with the flavor-changing
neutral current vertex $Z \bar t c$. In view of these constraints,
we discuss the experimental observability of the induced
rare decay mode $t \rightarrow Zc$,
both at the Fermilab Tevatron (with the Main Injector
or a luminosity-upgrade) and at the LHC.

\end{abstract}

\narrowtext

\section{Introduction}

The existence of the top quark has been established recently at
the Fermilab Tevatron
by the CDF and D0 collaborations  \cite{CDFD0}
with a measured mass  near 175 GeV.
Because the top quark
is heavier than all other observed fermions and gauge bosons and
has a mass of the order of the Fermi
scale,
it couples to the electroweak symmetry breaking sector strongly.
In its simplest incarnation,
the symmetry breaking sector of the Standard Model  (SM)
results from the presence of a complex
fundamental Higgs scalar
in the theory. However, there are theoretical arguments of
``triviality''  \cite{tri} and  ``naturalness'' \cite{natural} which
mitigate against  having such a simple scalar
sector.
One therefore believes
that the Higgs sector of the Standard Model is
just an effective theory, and  expects that
new physics phenomena will manifest
themselves  significantly through effective interactions of the
 top quark \cite{PZ}.

If anomalous top-quark couplings
beyond the SM were to exist,  they would
affect top-quark production and decay processes at hadron
and $e^+e^-$  colliders \cite{cmy,akr}.
Furthermore, such
couplings would also affect some quantities measured
with high precision,
 such as the partial width of
$Z \rightarrow b \bar b$.
The experimentally observed value at LEP-I
for the partial width ratio
$R_b = \Gamma(Z \rightarrow b \bar b) / \Gamma( Z \rightarrow {\rm Hadrons} )$
is somewhat higher than the Standard Model expectation \cite{LEP}.
This could be the result of a statistical fluctuation,  or  it could also be
an indication of new physics.

In this paper, we will study  the experimental constraints on
an anomalous top-quark coupling $Z \bar t c$\cite{PPZ} and discuss in detail
the  experimental observability of the induced rare decay mode
$t \rightarrow Z c$ \cite{fritzsch},
both at the Fermilab Tevatron and the LHC.
There are two major motivations for such a study.
First of all,  it is argued \cite{PPZ} that, in a dynamical electroweak
symmetry breaking scheme, the coupling
parameters,   $\kappa_{i j}$, which typify the
anomalous couplings of the $Z$ boson to fermions $i$ and  $j$,
may behave like,
\begin{equation}
\kappa_{i j} \sim O (\frac{ \sqrt{m_i m_j}}{v} ) ,
\end{equation}
with $m_i$ being the mass of the fermion $i$, and
$v$ the vacuum expectation value $v \simeq 250$ GeV.
This estimate is consistent with the bounds on the
 $\kappa_{i j}^{}$ for the light
quarks \cite{PPZ} and may give significant effects for the heavy top-quark
sector. Representative
models in which the above enhancement may be realized include
those with new dynamical interactions of the top quarks \cite{hill},
with multi-Higgs doublets \cite{multihiggs,htc},
and with new exotic fermions \cite{gronau}.
Secondly, the decay $t \rightarrow Z c$ is rather unique. It has a very
distinct experimental signature, especially if we look for the clean final
state from $Z \rightarrow l^+l^-$ at hadron colliders.
Moreover,
what makes this study  very important is that
the expected value of the branching fraction
BR$(t \rightarrow Z c )$ in the SM is extremely small,
being of an order of  10$^{-13}$ \cite{multihiggs}.
Thus, the observation of such a top-quark decay mode would signal the existence
of new physics.
As we shall see,  the Fermilab Tevatron with an integrated
luminosity of 10 fb$^{-1}$ and the LHC with 100 fb$^{-1}$
will have good  potential to explore this
anomalous $Z \bar t c$ vertex.

This paper is organized as follows. In Section II, we examine the current
low energy
constraints on the anomalous $Z \bar t c$ coupling. In Section III, we
study the possibility of testing this anomalous coupling at
the Fermilab Tevatron and the LHC.
Finally, in Section IV, we summarize  our results.
The helicity amplitudes for $t \rightarrow Zc$ decay are presented
in  an appendix.

\section{ Low energy constraints on the anomalous  $Z {\bar t}  c $
coupling   }

Following Ref. \cite{PZ}, we introduce
an effective lagrangian involving the anomalous top-quark couplings,
\begin{equation}
{\cal L}^{eff} = {\cal L}^{SM} + \Delta {\cal L}^{eff}
\,
\end{equation}
where ${\cal L}^{SM}$ is the Standard Model lagrangian and $\Delta {\cal
L}^{eff}$ includes all of the anomalous top-quark
couplings. For the purpose of this
paper, we consider only the anomalous couplings associated with the
$Z  \bar t c$ vertex, and
 only the operators of lowest dimension that contribute. Then
\begin{eqnarray}
\Delta {\cal L}^{eff} =   - \frac{g }{ {2 \cos\theta_W} } \,
[ \kappa_L^{} Z^\mu  \bar t \gamma_\mu (\frac{ 1 - \gamma_5}{2 } ) c
+\kappa_R^{} Z^\mu \bar t \gamma_\mu (\frac{ 1 + \gamma_5}{2} ) c ]
+ {\it h.c.} \ ,
\label{Leff}
\end{eqnarray}
where $g$ is the coupling constant of $SU(2)_L$,
$\theta_W$ is the Weinberg mixing angle and $\kappa^{}_{L(R)}$ are free
parameters determining the strength of these anomalous couplings.
Assuming $CP$-invariance,  $\kappa^{}_{L(R)}$ are real.

The lagrangian $\Delta {\cal L}^{eff}$ of  Eq. (\ref{Leff})
is written in the unitary gauge with
a non-linear realization of $SU(2)_L \times U(1)_Y$. The details of the
fermion transformation rule and
the procedure for constructing the invariant effective lagrangian
under the
nonlinear $SU(2)_L \times U(1)_Y$  group has been given in \cite{PZ}, and
will not be repeated here.
 However,
it is important
to emphasize that because $SU(2)_L \times U(1)_Y$ is spontaneously
broken down to $U(1)_{em}$, the anomalous  $Z\bar t c$ couplings, as well
as those of $Z\bar t t$ and $W\bar t b$,
can exist in dimension four, in contrast to the
anomalous couplings of
a photon  (or a gluon)  to $\bar t c$ or
$\bar t t$ which require the presence of the gauge field strengths
$F_{\mu \nu}$ ($G^a_{\mu \nu}$).

\subsection{Constraints on $\kappa_L$ }

The anomalous coupling $\kappa_{L}$ in Eq. (\ref{Leff}) is constrained
by experimental upper bounds on the induced flavor-changing neutral
couplings
of the light fermions.  Integrating the heavy top quark out of
${\cal L}^{eff}$ generates an effective
interaction of the form
\begin{equation}
\tilde{\cal L} = \frac{g}{\cos\theta_W} a_{i j} \bar {f_i} \gamma^\mu
              ( \frac{1 - \gamma_5}{2} ) f_j Z_\mu +{\it h.c.} \,
\end{equation}
where $f_i = b, \ s, \ d$. Evaluating the one-loop diagram for
the vertex correction
 gives
\begin{equation}
a_{ij} = \frac{\kappa_L^{}}{ 16 \pi^2 } \frac{m_t^2}{v^2} \
( V_{ti}V^*_{cj} + V_{tj}V^*_{ci} ) \ {\rm ln} \frac{\Lambda^2}{m_t^2},
\label{aij}
\end{equation}
where $V_{ij}$ are the elements of the Cabbibo-Kobayashi-Maskawa
matrix and
$\Lambda$ is a cutoff for the effective lagrangian.
The effective lagrangian in Eqs. (2) and
(3) is not renormalizable in the common sense.
 In our calculation
we use dimensional regularization to regularize the divergent loop integration
\cite{PPZ}. Practically, we have also focused only
on certain ``log-enhanced" terms by replacing
$1 / \epsilon$ by $ {\rm ln}(\Lambda^2 / m_t^2)$ in the final results.
Note  that although we perform the calculation
in the unitary gauge,  it can be shown easily that our results for $a_{ij}$
are gauge-independent.

To get a more stringent upper limit on $\kappa_L$, we
take the maximal values of the $V_{ti}$ and
$V_{cj}$ matrix elements from the particle data book \cite{databook}.
Further
we use the upper bounds on $a_{ij}$ derived in Ref. \cite{london} by
studying
several flavor-changing
processes,
such as $K_L \rightarrow
\bar \mu \mu$,
the $K_L-K_S$ mass difference, $B^0 - \bar {B^0}$ mixing and
$B \not \rightarrow l^+ l^- X$, reproduced below

$$a_{sd} < 2 \times 10^{-5}; \,  \, \, a_{bd} < 4 \times 10^{-4};
\,  \, \, a_{bs} < 2 \times 10^{-3}.$$
Taking
$\Lambda =$1 TeV
and
$m_t =$ 175 GeV, this gives an upper limit
\begin{equation}
\kappa_L < 0.05 .
\label{kllimit}
\end{equation}
This upper bound on $\kappa_L$
is more stringent than what can be derived from
the experimental measurement of the partial width of
$Z \rightarrow \bar b  b$ at  LEP-I.
That is, $\kappa_L$
would have to be bigger than $0.05$ to account for
the discrepancy with the SM expectations.
The bound obtained above using Eq. (\ref{aij}) depends on $\Lambda$,
so it only gives an order of
magnitude
 upper limit estimate on $\kappa_{L}$. Obviously these
bounds are not a substitute
for direct measurements.
However, perhaps surprisingly
 these experimental bounds are
close to the theoretical
values inferred for the
 anomalous $Z\bar t c$
couplings using Eq. (1).

\subsection{Constraints on $\kappa_R^{}$ }

For massless external
fermions, $\kappa_R^{}$ will not contribute to $a_{i j}$.
However, both $\kappa_R^{}$ and
$\kappa_L^{}$ contribute to the
oblique
corrections measured experimentaly at LEP-I and the SLC.
Consider  the vacuum
polarization tensor  \cite{PPZ}
$\Pi^{\mu\nu}_{ZZ} (q^2)$. The presence of
$\kappa^{}_{L(R)}$ will generate
 a term like
\begin{equation}
g^{\mu\nu} (-i) \frac{3}{16 \pi^2} \, \frac{g^2}{4 \cos\theta_W^2} \,
(\kappa_L^2 + \kappa_R^2) \,
(\frac{4}{3} q^2 - 2 m_t^2 ) \, {\rm ln} \frac{\Lambda^2}{m_t^2} .
\end{equation}
Such a contribution will affect
quantities like the $\rho$
and $S$ parameters.
The strongest constraint on $\kappa_R^{}$ comes from the $\rho$ parameter.
Using
the recent determination
of the change of $\rho$ from unity,
$\Delta \rho = 0.0004 \pm 0.0022 \pm 0.002$ \cite{databook},
one obtains a  $2\sigma$-limit,
\begin{equation}
\kappa^2_{tc} \equiv \kappa_L^2 + \kappa_R^2 \leq 0.084 .
\end{equation}
In view of the bound on $\kappa_L^{}$ of Eq. (\ref{kllimit}),
this gives a limit $\kappa_R \leq 0.29$.
The $S$ parameter does not improve on this
 upper bound for $\kappa_R$. For
$\kappa_L \sim 0.05 \, , \, \kappa_R \sim 0.29$, one finds that
$S^{new} \sim - 0.1$. This correction is safely
within the present experimental limit \cite{databook} for
$S^{new} = -0.42 \pm 0.36 ^{-0.08}_{+0.17}$.

We conclude this section by summarizing
the current upper bounds on $\kappa_L^{}$ and $\kappa_R^{}$:
\begin{equation}
\kappa_L^{} \leq 0.05; \quad
\kappa_R^{} \leq 0.29;  \quad {\rm or}  \quad \kappa_{tc}^{}\leq 0.29.
\end{equation}

\section{ Top-quark  Decay to $Z$-charm
at Hadron Colliders }

At the Fermilab Tevatron, the cross section for $t \bar t$ production
is about 5 pb with $\sqrt s = 2$ TeV \cite{xection}.
It is conceivable  that
a comprehensive study of top-quark
physics will be carried out
with the Main Injector (an integrated luminosity
about 1 fb$^{-1}$/yr expected),
or an upgraded Tevatron (about 10 fb$^{-1}$/yr).
The top quarks will be more copiously produced at the
LHC due to a much larger center of mass energy  (14 TeV)
and higher luminosity  (100 fb$^{-1}$/yr).
It is therefore interesting to
study the feasibility of testing  the
anomalous  $Z \bar t c$ couplings
 at these  hadron colliders.

\subsection{Top-Quark Decay to $Zc$}

To a good approximation, we can safely ignore the masses of
$c$ and $b$ quarks compared to $m_t$.
The decay $t \rightarrow Zc$  \cite{fritzsch} is similar to that
of  $t \rightarrow Wb$, up to the new couplings and a kinematical
correction factor. We can thus calculate the branching  fraction for
$t \rightarrow Zc$ by  the following simple formula,
\begin{equation}
BR(t\rightarrow Zc) \equiv
 \frac{\Gamma(t \rightarrow Zc)}{\Gamma(t \rightarrow Wb)}
= \frac{ \kappa_{tc}^2}{2}  \frac{(m_t^2 - M_Z^2)^2}{(m_t^2 - M_W^2)^2}
 \frac{(m_t^2 +2 M_Z^2)}{(m_t^2 + 2 M_W^2)}
\simeq 0.5  \kappa_{tc}^2.
\end{equation}
Figure~\ref{br} shows this branching fraction
plotted versus
$\kappa_{tc}^{} = \sqrt {\kappa_{L}^{2}+\kappa_{R}^{2}}$
for $m_t=$160 and 200 GeV.
{}From the figure  we see that the resulting
branching fraction is rather insensitive to the precise value of the
top-quark mass, so we will  take  $m_t=$175 GeV throughout the paper.
The $t \rightarrow Z c$ branching fraction is of order
0.1\% for  $\kappa_{tc}^{} \sim 0.05$, and  of order
1\% for  $\kappa_{tc}^{} \sim 0.15$.

The helicity amplitudes for the decay
$t \rightarrow Z c$  are  presented  in  an appendix.
One sees that a heavy top-quark  decays more significantly to
a longitudinally polarized vector boson.
Keeping only these terms,
the  charm-jet angular distribution from  a left-handed (right-handed)
top-quark decay  goes  approximately like
\begin{equation}
 \frac {d\Gamma_{L(R)}}{d\cos \theta} \sim \frac{m^2_t}{M^2_Z}
\, \bigl[\kappa^{2}_{L(R)} \, \cos^2 \frac{\theta}{2}+
\kappa^{2}_{R(L)} \, \sin^2 \frac{\theta}{2} \, \bigr].
\end{equation}
Here the top-quark polarization
and the charm-jet polar angle $\theta$
are defined in the top-quark rest frame
with respect to the direction of its motion.
One could imagine trying to explore the underlying couplings
$\kappa^{}_{L(R)}$
by studying the angular distributions of the decay products from
polarized top quarks. Unfortunately, the dominant  mechanisms
for top-quark production  at hadron colliders via $q \bar q$ and $gg$
annihilation
give little polarization. This
 makes the detailed study of these angular distributions
impossible.
To pursue this further,
one will have to study other production mechanisms with
larger top-quark polarization, such as
single-top production \cite{singlet,singlet2}. In  what follows,
we will ignore top-quark polarization effects altogether.

\subsection{Searching for a $t \rightarrow Zc$  Signal at  Hadron Colliders}

To obtain the signal event rates,
we  calculate the top-quark production via $q\bar q, gg \rightarrow t \bar t$
with lowest order matrix elements, but normalize the total cross sections
to values
which include higher order corrections \cite{xection}.
We have used the recent parton distribution functions MRS Set-A
\cite{mrsa}.
Due to the enormous  QCD backgrounds at hadron colliders,
it is very difficulty, if not impossible,
to search for the signal via the hadronic $W,Z$ decay channels, such as
$t \bar t \rightarrow W^\pm \, b Z c \rightarrow W^\pm +4$-jets
\cite{walter} or $Z +4$-jets \cite{barger}.
For our analysis, we therefore will mainly
concentrate on the pure leptonic decays of $W$ and $Z$, namely:
 $t \rightarrow Wb \rightarrow l^\pm \nu b$
and   $t \rightarrow Zc \rightarrow l^+ l^- c$,
calculated with exact decay matrix elements for on-shell $W,Z$.
We have ignored the spin
correlations for the decaying top quarks, as discussed above,
due to rather insignificant top-quark polarization \cite{bargeretal}
for the production mechanisms considered here.

Figure \ref{xsectn}  presents  the calculated total cross sections, plotted
versus the anomalous coupling $\kappa_{tc}$, for the process
$t \bar t \rightarrow W^\pm \, b Z c \rightarrow l^\pm \nu, l^\pm l^\mp jj$,
where $l=e, \mu$ and $j$ denotes a jet from a $b$ or $c$ quark. We see
that  for  $\kappa_{tc}$  in the range of interest, the signal cross section
at the Tevatron is rather small, of the order of a few  fb,
while at LHC energies, it is about two orders of magnitude larger.
However, the final state of the events is  quite distinct:
three isolated charged leptons, two of which reconstruct a $Z$, large
missing transverse energy ($E^{miss}_T$), and two hard jets coming
from the $b$ or $c$ quarks.
The only irreducible background to this signal  is the electroweak
process $p \bar p, pp \rightarrow W^\pm Z X\rightarrow  l^\pm \nu l^+ l^- X$.
Since the signal events naturally contain two energetic jets from heavy
top-quark decays, typically with a transverse momentum of order
$p^{}_T(j) \simeq  \frac{1}{2} m_t \, (1-M_W^2/m_t^2)$, it is advantageous
to demand two observable jets in the events to suppress the
$WZ$ background.

To better address the feasibility of observing the effects from $\kappa_{tc}$,
we need to consider detector effects and to
impose some acceptance cuts on the transverse momentum $(p_T^{})$,
pseudo-rapidity $(\eta)$, and the separation in azimuthal angle-pseudo rapidity
plane $(\Delta R)$ for the charged leptons and jets.  We choose
 for the Tevatron the acceptance cuts:
\begin{eqnarray}
p_T^l &>&15 \; {\rm GeV}, \quad  |\eta^l| < 2.5, \quad  \Delta R_{lj} > 0.4,
\quad
E^{miss}_T>20 \; {\rm GeV},  \nonumber \\
p_T^j&>&15 \; {\rm GeV}, \quad  \quad  |\eta^j| < 2.5, \quad  \Delta R_{jj} >
0.4.
\label{tevcut}
\end{eqnarray}
For the LHC the equivalent cuts chosen are:
\begin{eqnarray}
p_T^l &>&25 \; {\rm GeV}, \quad  |\eta^l| < 3, \quad  \Delta R_{lj} > 0.4,
\quad
E^{miss}_T>30 \; {\rm GeV},  \nonumber \\
p_T^j&>&30 \; {\rm GeV}, \quad  \quad  |\eta^j| < 3, \quad  \Delta R_{jj} >
0.4.
\label{lhccut}
\end{eqnarray}
With these minimal acceptance cuts, the signal and background rates at
the Tevatron and LHC are given in Fig.~\ref{rate} (dotted curves).

 There
is an easy way to improve the signal-to-background ratio to some extent.
Recall the argument
given above that the jet $p^{}_T(j)$ spectrum  for the signal
peaks around $m_t/2$.
In contrast,
the
QCD jets in the background events are near the threshold region (determined
by the cutoff imposed) and extend further, if phase space is available as
it is
at LHC energies.  This feature is demonstrated in Fig.~\ref{ptjj},
where the differential cross section versus a scalar sum of the
jet's transverse momenta,
$$
p^{}_T(jj) \equiv |\vec p^{}_T(j_1)|  + |\vec p^{}_T(j_2)|,
$$
is plotted,  where the anomalous coupling $\kappa^{}_{tc}$
has been chosen to be 0.14, for which the signal and the background
have about the same total rate.
We see indeed, that the $p^{}_T(jj)$ spectrum for the signal
peaks near $m_t$ and does not significantly depend on the c.m. energy.
If we further impose a cut on  $p^{}_T(jj)$,
\begin{equation}
80 \, \, {\rm GeV}  < p^{}_T(jj) < 250 \, \, {\rm GeV},
\label{ptjjcut}
\end{equation}
the background is reduced,
but the signal is hardly
affected. The effect of this cut
is shown in Fig.~\ref{rate}  by the dashed curves.

An important advantage of searching for
top-quark flavor-changing decays in the $t\rightarrow Zc \rightarrow l^+ l^- c$
channel is the full reconstructability of the top-quark mass.
The mass variable  $M(l^+l^-j)$  from the decay
$t \rightarrow Zc \rightarrow l^+l^-j$ should reconstruct   $m_t$,
and is clearly the most characteristic quantity for the signal.
Similarly, the  mass variable $M(l \nu j)$ from the decay
$t \rightarrow Wb \rightarrow l \nu j$ should also peak near $m_t$.
However, in
the latter case there is a two-fold ambiguity in constructing the neutrino
momentum  along the beam direction \cite{pnu} due to the lack of
knowledge of the parton c.m. frame.
For an input $M_W^{}$ and massless
leptons,  using the measured charged lepton momentum ($p^\ell$)
and the {\it transverse} momentum of the neutrino ($p_T^\nu$),
the two solutions for the {\it longitudinal} momentum of the neutrino
are given by
\begin{eqnarray}
\noalign{\vskip 5pt}
p_L^{\nu} = {1\over 2\, (p_T^{\ell})^2 } \Biggl\{ p_L^{\ell}
\Bigl(M_W^2 + 2\, \hbox{\bf p}_T^{\ell} \cdot \hbox{\bf p}_T^{\nu} \Bigr)
\pm \, p^\ell \, \biggl[ \Bigl(M_W^2 + 2\, \hbox{\bf p}_T^{\ell} \cdot
\hbox{\bf p}_T^{\nu} \Bigr)^2
- 4\, (p_T^\ell)^2\, (p_T^\nu)^2 \biggr]^{1/2} \Biggr\} \>.
\label{plnu}
\end{eqnarray}

Although the
intrinsic top-quark width is only about 1.5 GeV for $m_t=175$ GeV, the
signal peaks for these mass variables will not be as narrow and sharp,
due to the finite detector  resolution for measuring
 the energy
of the charged leptons and hadrons. We simulate these
 detector effects by assuming
a Gaussian energy smearing for the electromagnetic and hadronic
calorimetry as follows:
\begin{eqnarray}
\Delta E/E &=& 30\%/\sqrt E \oplus 1\%, \quad  {\rm for \ \ leptons}  \nonumber
\\
                   &=&  80\%/\sqrt E \oplus 5\%, \quad  {\rm for \ \  jets.}
\label{smear}
\end{eqnarray}
Moreover,
in constructing $M(Zc)=M(l^+ l^- j)$,  because we cannot straightforwardly
identify the $c$-quark jet, we examine both jets in the
final state and keep the  one of the two
mass values  which is numerically closer
to $m_t=175$ GeV.  Figure \ref{mllj} shows the differential cross section
for the reconstructed mass
variable $M(l^+l^-j)$ obtained by this procedure
for the signal (solid histogram) and
background (dashed histogram) at the Tevatron a) and  the LHC b).
We see that the
signal has a sharp distribution near $m_t$  with
a width of
about 20 GeV, dominated by the hadronic calorimeter resolution of
Eq.~(\ref{smear}).
  In this figure,  the anomalous coupling $\kappa^{}_{tc}$
has been chosen to be 0.1, a value for which the signal and the background
have about the same total rate.
One can further study the mass variable
$M(W b) =  M(l \nu j)$.
Here we once again make use of the
knowledge of the value of $m_t$ and choose from
the two $M(l \nu j)$ values the one
which is numerically closer to $m_t$.
Figure \ref{mlnuj} presents the differential cross section
for the variable $M(l^\pm \nu j)$ for the signal (solid histogram) and
background (dashed histogram) for the
Tevatron and the LHC.  We see again
a sharp distribution near $m_t$  with  a width now  about 30 GeV.

Making use of
the distributions of Figs.~\ref{mllj} and
\ref{mlnuj},
which are rather impressive,
 allows us to estimate the statistical sensitivity with which
one can hope to measure
$\kappa_{tc}$. This can be carried out by searching for events which fall
within the ranges
\begin{equation}
|M(l^+l^-j) - m_t \,| < 20 \, {\rm GeV}, \quad
|M(l^\pm \nu j)-  m_t \,| < 40 \, {\rm GeV}.
\label{dmtop}
\end{equation}
The solid curves in Fig.~\ref{rate} show the signal and background rates
with all cuts in Eqs.~(\ref{tevcut}),
 (\ref{lhccut}), (\ref{ptjjcut}) and (\ref{dmtop}).
Note again that the cuts  in Eq.~(\ref{dmtop})
 do  little to diminish the signal.
 Figure.~\ref{lum} gives the  sensitivity with
which  $\kappa_{tc}$ can be measured as a function of
 accumulated luminosity in units of
fb$^{-1}$ at the Tevatron  a) and the LHC b)  for a 99\% Confidence Level
(C.L.)
(solid)  and a 95\% C.L. (dashed).
Poisson statistics has been adopted in determining the Confidence Level
when the number of events is small. For a large number of events,
a 95\%  (99\%) C.L.  in this scheme approximately
corresponds to Gaussian statistics with
$\sigma \equiv  N_S/\sqrt{N_S+N_B} \simeq 3 (4)$.

In probing $\kappa^{}_{tc}$  at a 99\% C.L.,
we find  the sensitivity  to be, at the Tevatron
\begin{equation}
\begin{array}{lcc}
{\rm with} \; 1 \; {\rm fb^{-1}}:   \quad   & \kappa^{}_{tc} \sim 0.42 \quad
{\rm or}
& BR(t \rightarrow Zc) \sim 9\% \\
{\rm with} \; 3 \; {\rm fb^{-1}}:   \quad   & \kappa^{}_{tc} \sim 0.26 \quad
{\rm or}
& BR(t \rightarrow Zc) \sim 3\% \\
{\rm with} \;10 \; {\rm fb^{-1}}:  \quad    & \kappa^{}_{tc} \sim 0.16 \quad
{\rm or}
& BR(t \rightarrow Zc) \sim 1\% \\
{\rm with} \; 30 \; {\rm fb^{-1}}: \quad      & \kappa^{}_{tc} \sim 0.11 \quad
{\rm or}
& BR(t \rightarrow Zc) \sim 0.6\% \, ,
 \end{array}
\end{equation}
and at the LHC
\begin{equation}
\begin{array}{lcc}
{\rm with} \; 1 \; {\rm fb^{-1}}:   \quad   & \kappa^{}_{tc} \sim 9\cdot
10^{-2}
\quad {\rm or}  & BR(t \rightarrow Zc) \sim 4\cdot 10^{-3} \\
{\rm with} \;10 \; {\rm fb^{-1}}:  \quad    & \kappa^{}_{tc} \sim 4\cdot
10^{-2}
\quad {\rm or} & BR(t \rightarrow Zc) \sim 8\cdot 10^{-4} \\
{\rm with} \; 100 \; {\rm fb^{-1}}: \quad      & \kappa^{}_{tc} \sim 2\cdot
10^{-2}
\quad {\rm or} & BR(t \rightarrow Zc) \sim 2\cdot 10^{-4} \, .
\end{array}
\end{equation}
We see that a high luminosity Tevatron or the LHC would have
good potential to explore this anomalous $Z\bar t c$ coupling.

\section{Discussions and Summary}

We have thus far only concentrated on the leptonic decays
$Z \rightarrow l^+ l^- $.
Given the possibility of  efficient $b$-tagging at the Tevatron \cite{CDFD0}
and LHC \cite{atlascms}
detectors,  one may consider also studying the decay $Z \rightarrow
b \bar b$, with a branching fraction of about 15\%. The signal rate would
be approximately doubled
and this mode might prove quite interesting if one could keep
  the QCD background under
control with $b$-tagging.

Although not addressed in this paper, it may be equally possible that
there exist other effective operators resulting in flavor-changing neutral
current interactions, such as $g \bar t c$
 and $\gamma \bar t c$.
These operators may have different physics origin and should be studied
separately. In a hadron collider environment,
the signals associated with these vertices may not be
easily separated from the QCD backgrounds. This issue is currently
under investigation \cite{young}.

To summarize,  we have examined the low
energy experimental constraints on a possible
anomalous top-quark coupling associated with the flavor-changing
neutral current vertex $Z \bar t c$. The  limits derived from the current
experimental data are
\begin{equation}
\kappa_L^{} \leq 0.05; \quad \kappa_R^{} \leq 0.29.
\end{equation}
We  have studied the experimental observability of the induced
rare decay mode $t \rightarrow Zc$,  both at  the
Fermilab Tevatron and at the LHC.  At a 99\% C.L.,
a three-year running of the Tevatron Main Injector (3 fb$^{-1}$)
would reach a sensitivity of BR$(t \rightarrow Zc) \simeq $3\%;
with an integrated luminosity of 10 fb$^{-1}$ at an upgraded Tevatron,
one should be able to explore new physics at a level of
BR$(t \rightarrow Zc) \simeq $1\%. At the LHC,
one can reach a value of $\kappa_{tc} \simeq 0.02$ or
BR$(t \rightarrow Zc) \simeq 2\cdot 10^{-4}$ with  a luminosity of
100  fb$^{-1}$.
Because of the cleanliness of the experimental signature and the
smallness of the SM prediction for this channel,
observation of this rare decay  would  clearly
signal new physics beyond the SM.
If seen, such a signal would provide important clues to
understanding the mechanism of electroweak symmetry breaking and
possibly fermion (especially top-quark) mass generation.

\section{Acknowledgments}

We would like to thank C. Hill, S. Parke, S. Willenbrock for discussions.
X.Z. would like to thank the Fermilab theory group for the hospitality
during the final stage of this work. This work was supported in part by
the U.S.~Department of Energy
under Contracts  DE-FG03-91ER40674 (T.H.), DE-FG03-91ER40662 (R.D.P.),
DE-FG02-94ER40817 (X.Z.).

\section*{Appendix: \, Helicity Amplitudes for  top to Z-charm}

In the limit of $m_c=0$,
the non-zero helicity amplitudes for  $t \rightarrow cZ$ decay, denoted by
 ${\cal M}(\lambda_t, \lambda_c, \lambda_Z^{})$,
after suppressing a common factor $[g/(2 \cos \theta_W)] \sqrt{2m_t E_c}$,
are

\begin{eqnarray}
{\cal M}(- - 0) &=& \frac{m_t}{M_Z} \kappa_L^{} \cos \frac{\theta}{2},
\nonumber  \\
{\cal M}(- - -) &=& \sqrt{2} \, \kappa_L^{} \sin \frac{\theta}{2} , \nonumber
\\
{\cal M}(- + 0) &=& \frac{m_t}{M_Z} \kappa_R^{} \sin
\frac{\theta}{2}e^{-i\phi}, \nonumber  \\
{\cal M}(- + +) &=& \sqrt{2} \, \kappa_R^{} \cos \frac{\theta}{2} e^{-i\phi},
\nonumber  \\
{\cal M}(+ - 0) &=& -\frac{m_t}{M_Z} \kappa_L^{} \sin
\frac{\theta}{2}e^{i\phi}, \nonumber \\
{\cal M}(+ - -) &=& \sqrt{2}  \kappa_L^{} \cos \frac{\theta}{2} e^{i\phi},
\nonumber \\
{\cal M}(+ + 0) &=& \frac{m_t}{M_Z} \kappa_R^{} \cos \frac{\theta}{2},
\nonumber\\
{\cal M}(+ + +) &=& -\sqrt{2} \, \kappa_R^{} \sin \frac{\theta}{2}.
\end{eqnarray}
Here  $\theta$ is the polar angle of the $c$ quark
in the top-quark rest frame, measured
with respect to the top-quark momentum direction; similarly,
the helicities ($\lambda$'s) have been defined in the top-quark rest frame
and with respect to the moving particle direction.

%
%
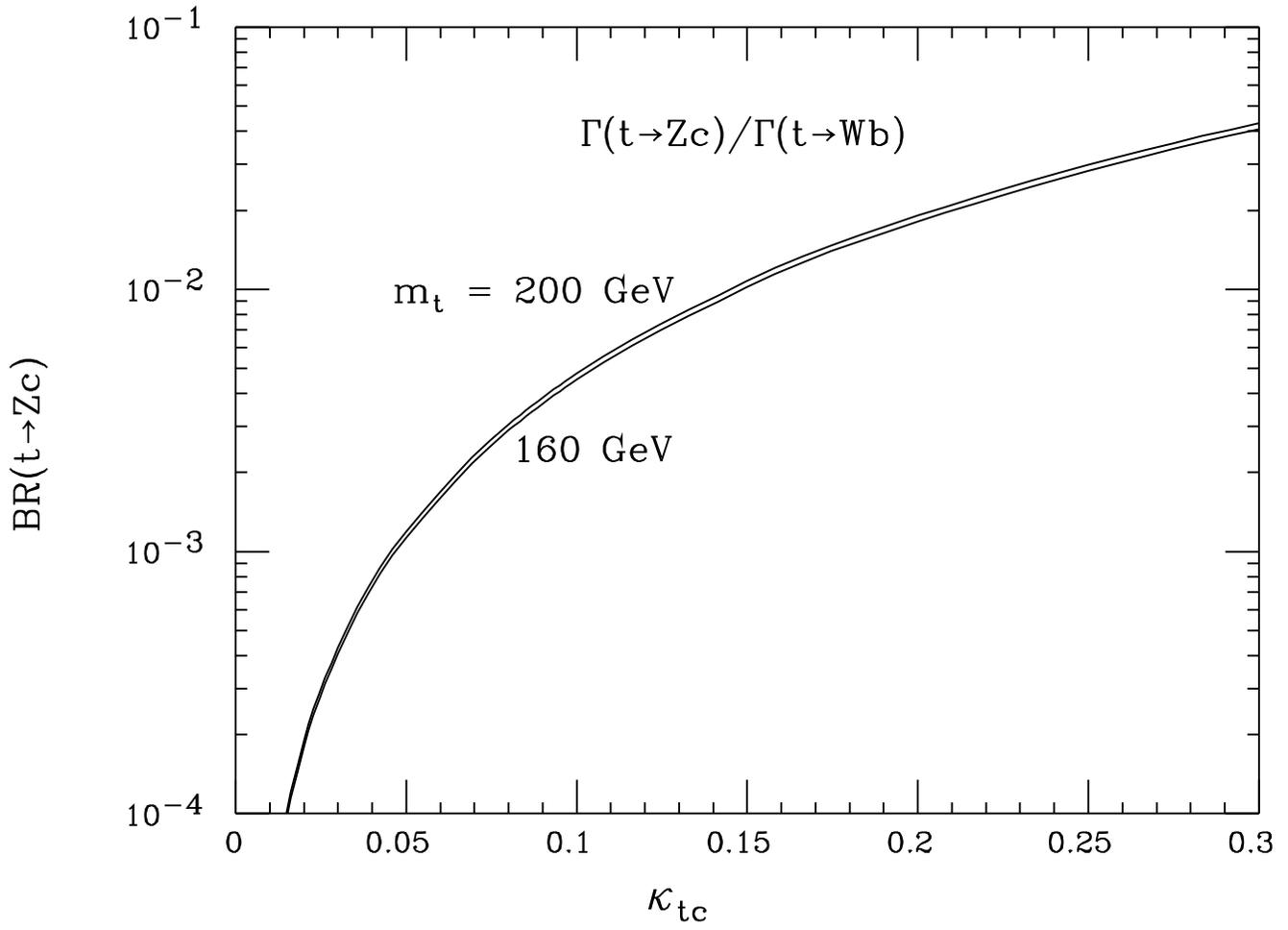
\begin{figure}[h]
\caption{\label{br}
Branching fraction   $\Gamma(t \rightarrow Zc)/\Gamma(t \rightarrow Wb)$
plotted
versus the anomalous  coupling $\kappa_{tc}^{}$ for $m_t=$160 and 200 GeV.
}
\end{figure}
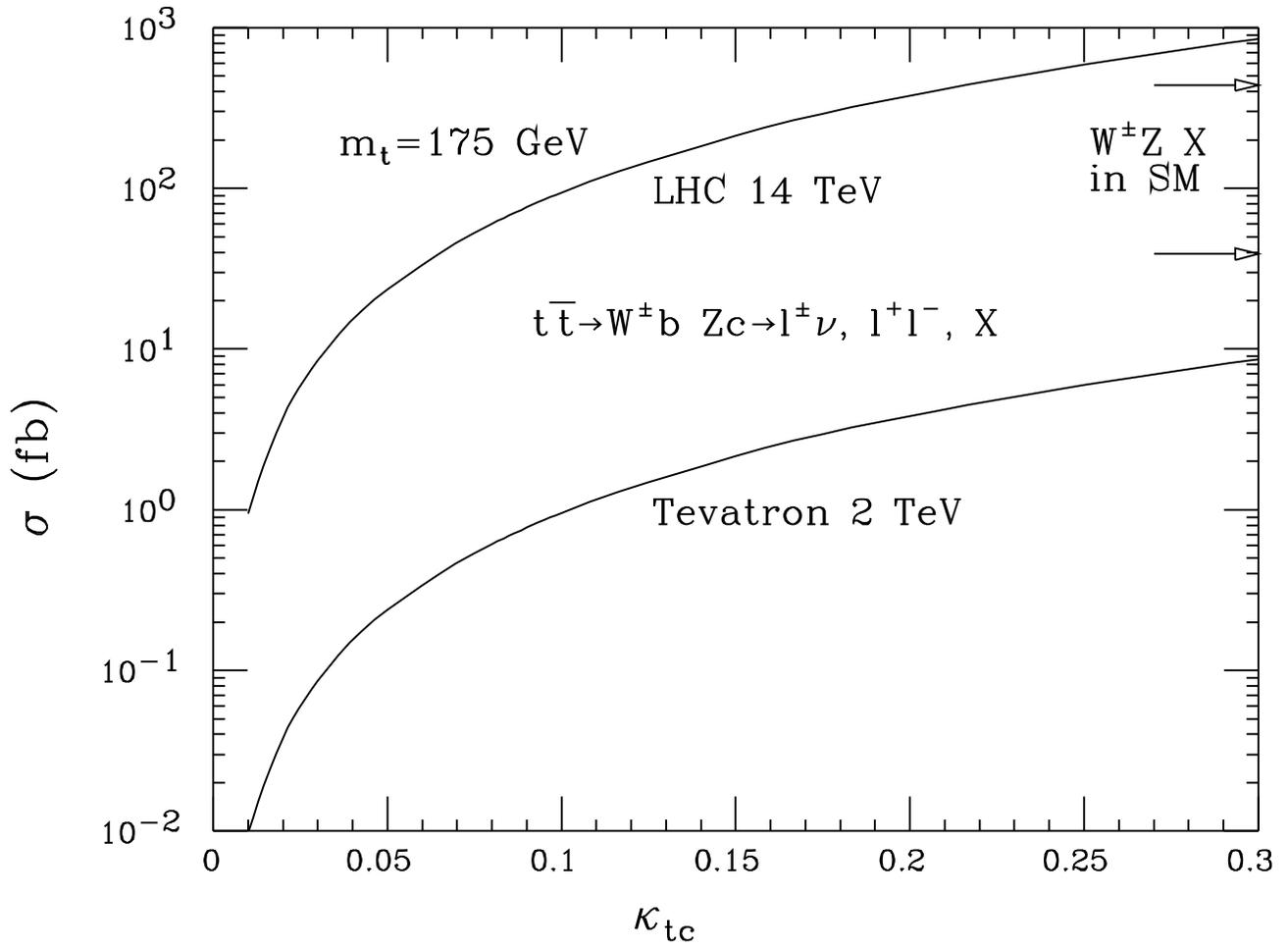
\begin{figure}[h]
\caption{\label{xsectn}
Total cross section for
$p \bar p, pp \rightarrow t \bar t X
\rightarrow W^\pm b Zc X \rightarrow l^\pm \nu l^+ l^- X$ in units of fb
for $m_t=175$ GeV, at  the Tevatron and LHC plotted
versus $\kappa^{}_{tc}$.
The arrows on the right-hand  scale indicate the inclusive rates
at the Tevatron (lower arrow) and  LHC (upper arrow) for
$ p \bar p, pp \rightarrow W^\pm Z X\rightarrow  l^\pm \nu l^+ l^-X$
production.}
\end{figure}
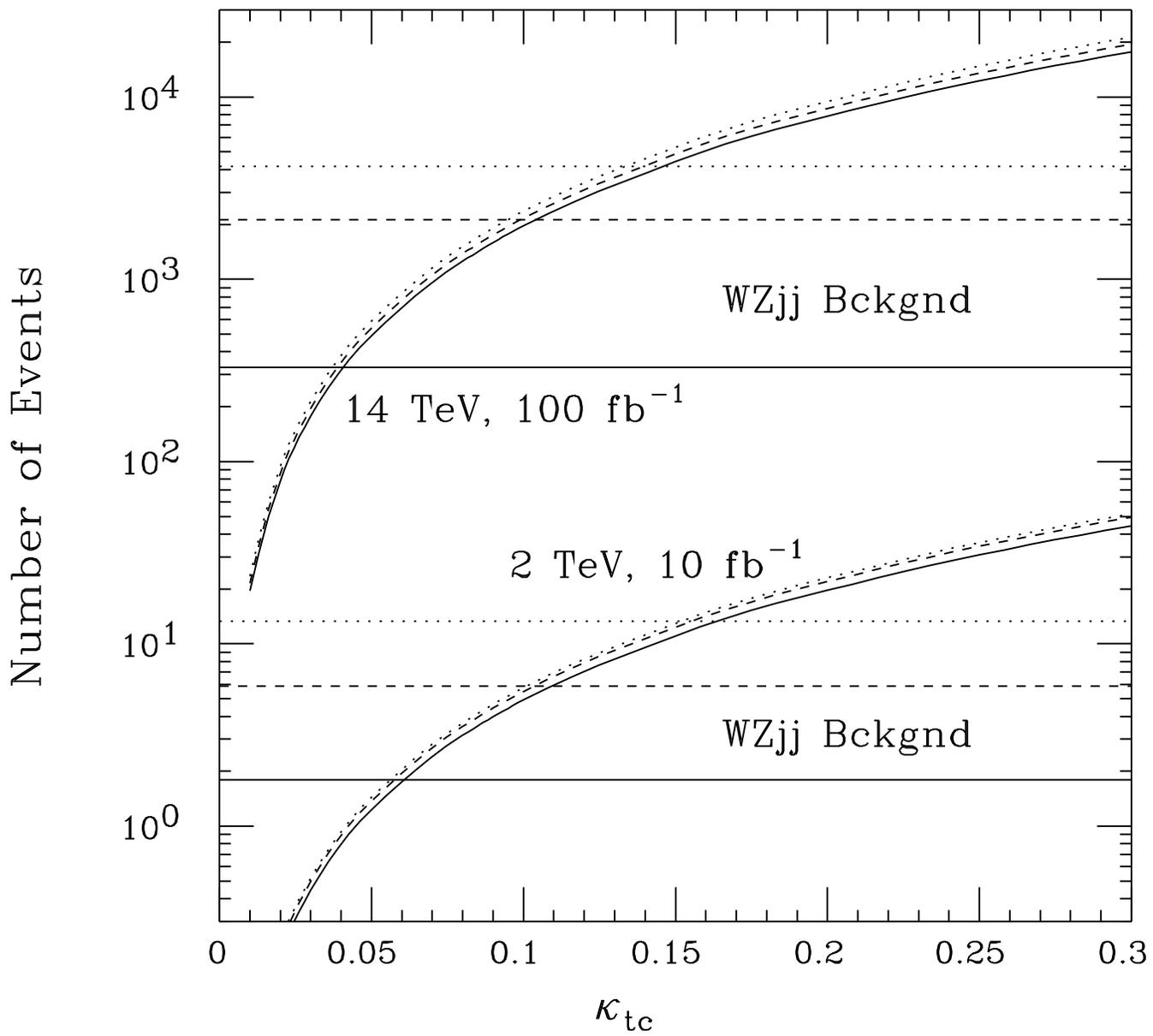
\begin{figure}[h]
\caption{\label{rate}
\narrower
Expected number of events  for $m_t=175$ GeV
at  the Tevatron with an integrated
luminosity of 10  fb$^{-1}$
and the
LHC with 100 fb$^{-1}$ plotted versus  $\kappa^{}_{tc}$. The horizontal lines
are for the $W^\pm Z jj$ background. The dotted curves are the rates
after imposing the minimal acceptance cuts
in  Eqs.~(\protect\ref{tevcut}) and (\protect\ref{lhccut});
the dashed
curves also include
the cuts in Eq.~(\protect\ref{ptjjcut});
the solid
curves include, in addition,
the mass cuts in Eq.~(\protect\ref{dmtop}).
A Gaussian energy smearing according to Eq.~(\protect\ref{smear}) is
included in the simulation. }
\end{figure}
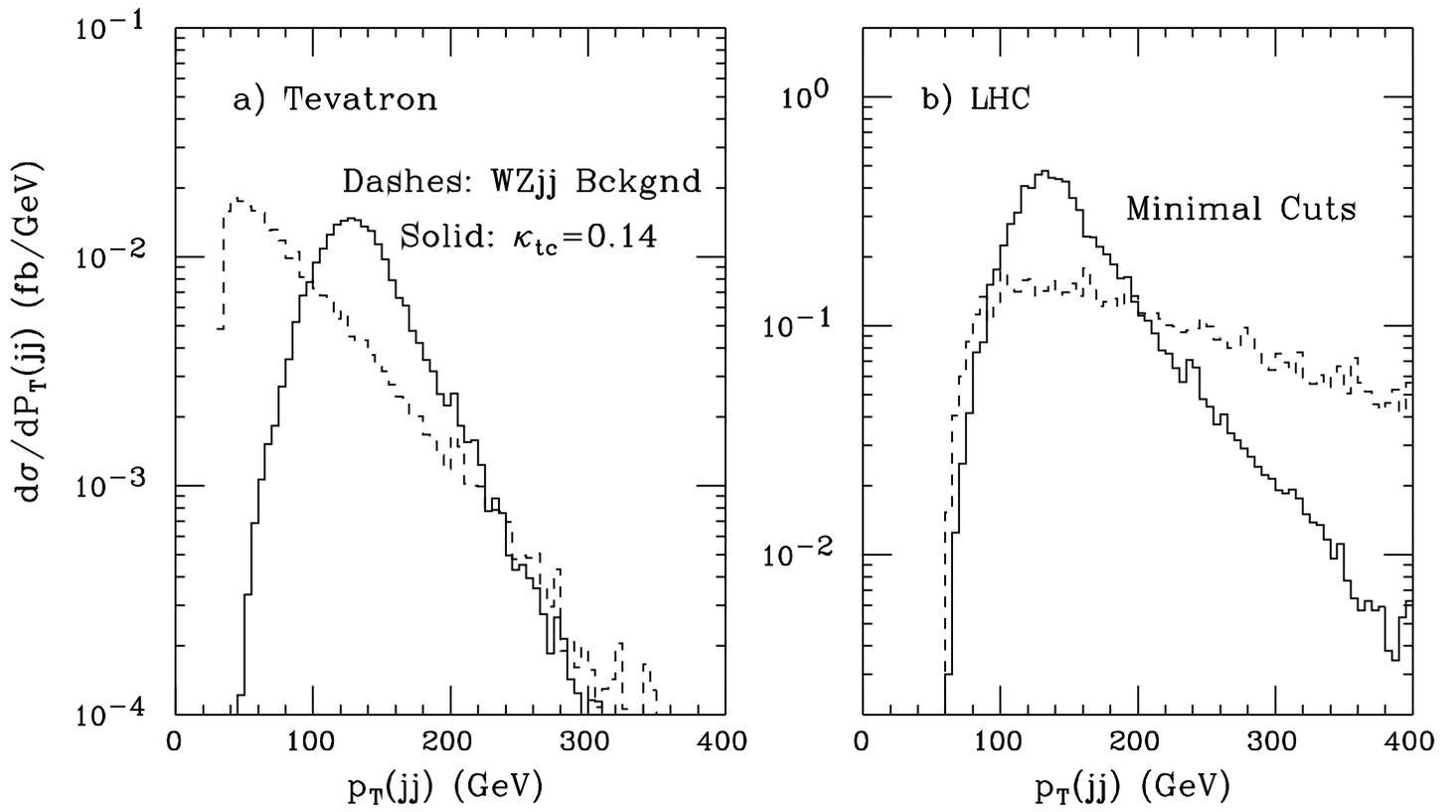
\begin{figure}[h]
\caption{\label{ptjj}
Differential cross sections $d\sigma/dp^{}_T(jj)$ in units of  fb/GeV
at a). the
Tevatron and b). the LHC
 with $m_t=175$ GeV and
$\kappa^{}_{tc}=0.14$ for the signal
(solid histogram)  and background (dashed histogram).
The minimal acceptance cuts
in  Eqs.~(\protect\ref{tevcut}) and (\protect\ref{lhccut})
have been imposed.
}
\end{figure}
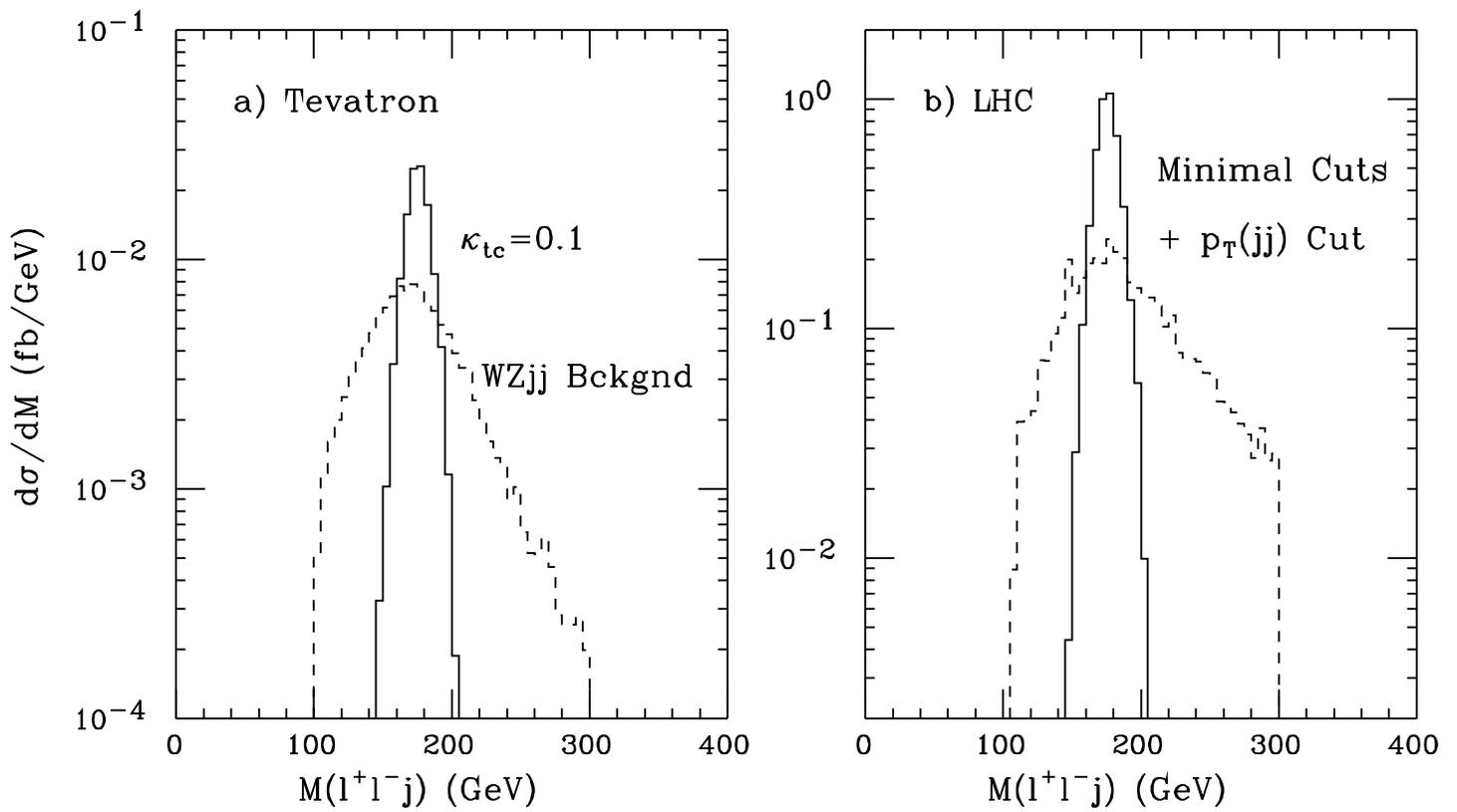
\begin{figure}[h]
\caption{\label{mllj}
Differential cross sections $d\sigma/dM(l^+l^-j)$ in units of  fb/GeV
at a). the Tevatron and b). the LHC
  with $m_t=175$ GeV
and $\kappa^{}_{tc}=0.1$ for the signal
(solid histogram)  and background (dashed histogram).
The acceptance cuts  in  Eqs.~(\protect\ref{tevcut}),
(\protect\ref{lhccut}) and  (\protect\ref{ptjjcut})  have been imposed.
}
\end{figure}
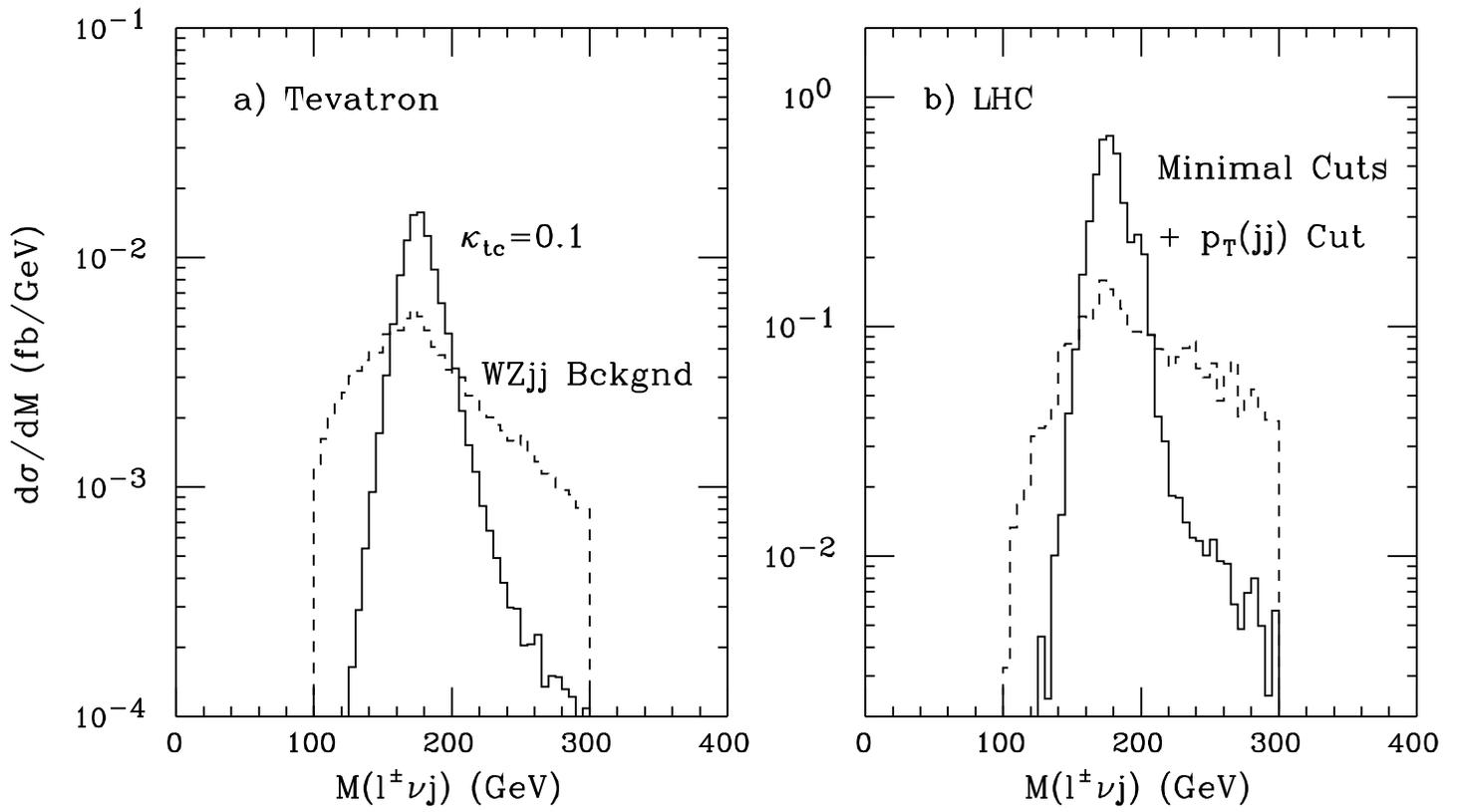
\begin{figure}[h]
\caption{\label{mlnuj}
Same as Fig. \protect\ref{mllj},  but for the reconstructed mass distributions
$d\sigma/dM(l^\pm \nu j)$.
}
\end{figure}
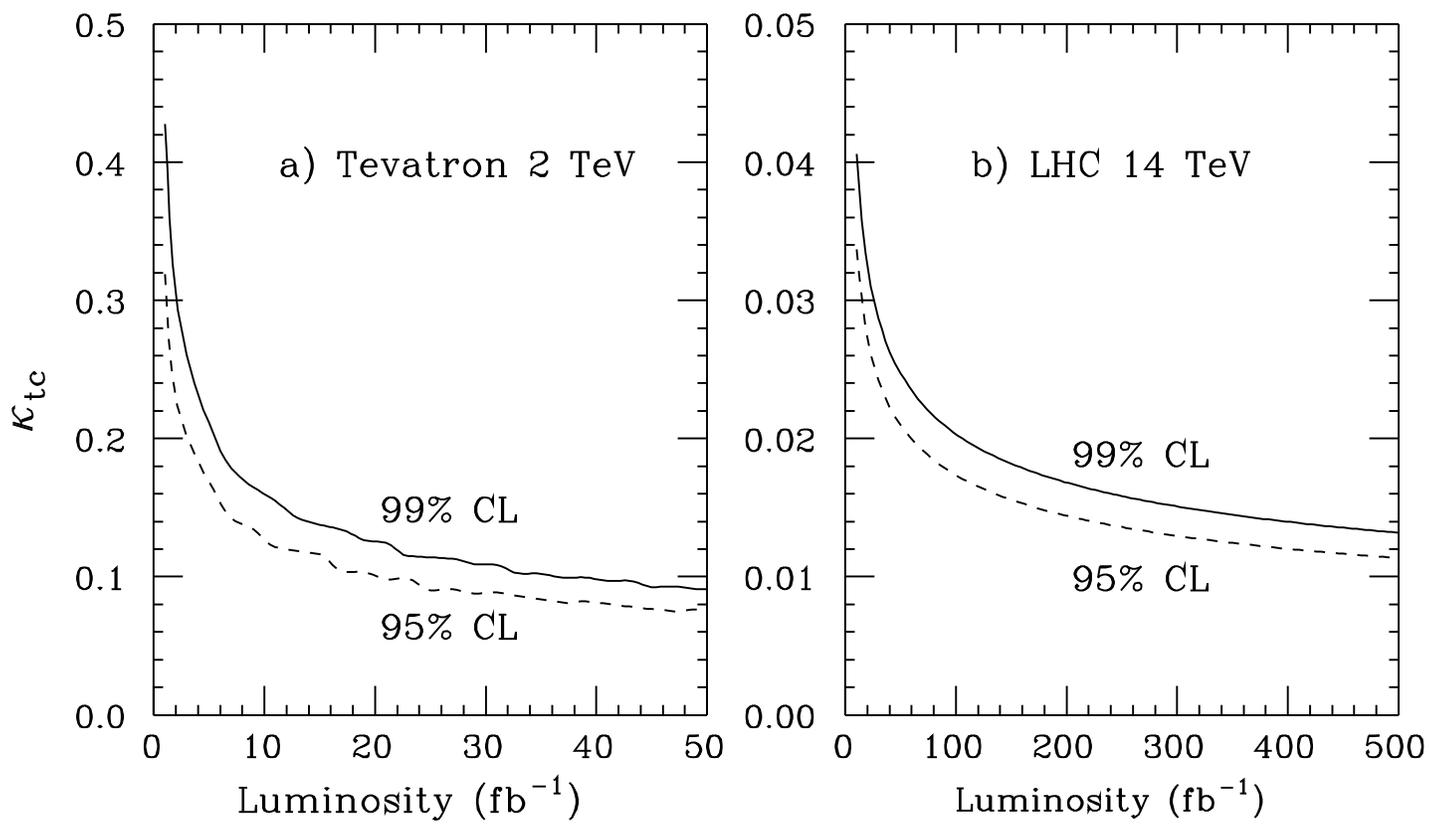
\begin{figure}[h]
\caption{\label{lum}
Sensitivity on $\kappa_{tc}$ for $m_t=175$ GeV
plotted versus the accumulated luminosity
(fb$^{-1}$) at a). the Tevatron and b). the LHC for a 99\% C.L.
(solid)  and a 95\% C.L. (dashed).
}
\end{figure}

\end{document}